\documentclass[10pt,conference]{IEEEtran}
\IEEEoverridecommandlockouts
\usepackage{cite}
\usepackage{amsmath,amssymb,amsfonts}
\usepackage{algorithmic}
\usepackage{graphicx}
\usepackage{textcomp}
\usepackage{xcolor}
\usepackage{multirow}
\usepackage{hyperref}
\usepackage{booktabs}
\usepackage{xspace}
\usepackage{subcaption}
\usepackage[flushleft]{threeparttable}

\usepackage{listings}
\colorlet{punct}{red!60!black}
\definecolor{background}{HTML}{EEEEEE}
\definecolor{delim}{RGB}{20,105,176}
\colorlet{numb}{magenta!60!black}

\lstdefinelanguage{json}{
basicstyle=\normalfont\ttfamily,
numbers=none,
numberstyle=\scriptsize,
stepnumber=1,
numbersep=8pt,
showstringspaces=false,
breaklines=true,
frame=lines,
backgroundcolor=\color{background},
}

\usepackage[linesnumbered]{algorithm2e}
\RestyleAlgo{ruled}
\SetKwComment{Comment}{/* }{ */}
\SetKwInOut{Input}{Input}
\SetKwInOut{Output}{Output}

\def\BibTeX{{\rm B\kern-.05em{\sc i\kern-.025em b}\kern-.08em
    T\kern-.1667em\lower.7ex\hbox{E}\kern-.125emX}}
\begin{document}

\newcommand{\myapp}{\textsc{MergeShuffle}\xspace}
\newcommand{\mymerge}{\textsc{Merge}\xspace}
\newcommand{\myshuffle}{\textsc{Shuffle}\xspace}
\newcommand{\linearshuffle}{\textsc{LinearShuffle}\xspace}
\newcommand{\tspshuffle}{\textsc{TspShuffle}\xspace}

\title{\myapp : Reconstruct Shared Libraries for Improved Performance and Security
}

\author{\IEEEauthorblockN{Huaifeng Zhang}
\IEEEauthorblockA{\textit{Chalmers University of Technology} \\
Gothenburg, Sweden \\
huaifeng@chalmer.se}
\and
\IEEEauthorblockN{Ahmed Ali-Eldin}
\IEEEauthorblockA{\textit{Chalmers University of Technology} \\
Gothenburg, Sweden \\
ahmed.hassan@chalmers.se}
}

\maketitle

\begin{abstract}
Debloating shared libraries has been an active research topic for years. Tools for shared library debloating rely on either static or dynamic analysis techniques to discover which parts of the shared library is used. Once these parts are found, these tools take one of two routes, either removing parts of the library that are not used, requiring reconstruction of the library; or they simply overwriting the unused functions. However, recently a broad study~\cite{brown2024broad} on debloating tools has shown that some of these tools either increase the static binary size, or are generally less effective for complex workloads.

In this paper, we argue that shared-library debloaters' performance can be improved by proper reconstruction of the shared-libraries once the used functions in the library are identifies. We propose \myapp, a novel tool for reconstructing shared libraries from a given set of file ranges to eliminate unused code based on user-specified requirements, without requiring access to source code.
\myapp takes as input a shared library and a set of file ranges to retain, and performs a two-phase reconstruction: merging nearby ranges to reduce loading overhead, and shuffling them to minimize file size while preserving page-aligned memory mappings.
The shuffling problem is formulated as a variant of the Traveling Salesman Problem and solved using a heuristic algorithm.
\myapp utilizes program headers to remap retained code to its original memory addresses, preserving user-specified functionality.
Our results show that \myapp achieves up to 60\% reduction in file size, 16\% reduction in memory usage, and 32\% reduction in runtime, while maintaining user-specified functionality and improving security.

\end{abstract}


\section{Introduction}

Modern software engineering relies on software reuse~\cite{narasimhan2015copy}.
At each phase, new components—code, dependencies, and resources—are introduced, while the removal of unused or obsolete elements is rarely prioritized~\cite{shackleton2023dead}.
Generally, during development, features are continuously added, sometimes leading to \emph{feature creep}~\cite{elliott2007anything}.
In the build phase, the source code is compiled and often linked with numerous third-party libraries~\cite{fan2020escaping}.
Finally, during deployment, applications are frequently containerized, bundling all dependencies and assets, regardless of whether they are necessary~\cite{zhang2024machine,zhao2020large}.

This pervasive addition leads to software bloat: the inclusion of unused code, features, and resources~\cite{hubert2024bloat}.
Bloated software consumes excessive disk space and memory and increases the system's attack surface, potentially introducing security vulnerabilities~\cite{zhang2024machine,porter2023decker}.
The problem is particularly acute for shared libraries.
Shared libraries are designed to accelerate application development by offering reusable components~\cite{gartner2021report}.
Any bloat or vulnerability in a library propagates to all applications that depend on it~\cite{shen2025understanding}.
Moreover, a fundamental gap exists between library developers and library users: developers aim to maximize adoption by incorporating many features, while individual library users typically require only a small subset of features.
This gap has further widened with the advent of micro-services and serverless computing.
Services are increasingly modular, narrow, and task-specific, yet they often rely on large, general-purpose shared libraries that are not tailored to their precise needs~\cite{liu2023faaslight,choi2019lambda}, leading to significant inefficiencies, especially during cold start scenarios~\cite{tariq2025efficient}.

Shared libraries — which are the most common form of third-party libraries — suffer significantly from this gap due to their widespread use in both operating systems and applications~\cite{quach2018dpiecewise,agadakos2019nibbler,ziegler2019honey}.
For instance, Quach et al. report that on Ubuntu Desktop, only about 5\% of the GNU C standard library (glibc) is used on average by the programs depending on it.
Even the heaviest user of glibc in their survey (the VLC media player) invoked merely 18\% of the library’s code~\cite{quach2018dpiecewise}.
These rarely used features not only bloat the shared library but also increase the attack surface, making it more vulnerable to exploits~\cite{agadakos2020large}.
For example, the Heartbleed vulnerability in OpenSSL was caused by a bug in a rarely-used TLS heartbeat feature – an optional extension that most users did not need, yet it was present in all builds and became a critical exploit vector~\cite{carvalho2014heartbleed}.
Furthermore, malicious code can be injected into shared libraries through supply chain attacks, as was recently demonstrated by the XZ backdoor incident~\cite{nistCVE20243094}.

Despite ubiquity of shared libraries in modern software ecosystems, there exists limited support for tailoring them to the specific needs of individual users.
This challenge has attracted attention from both academia and industry.
From academia, several software debloating tools have been proposed for shared libraries~\cite{quach2018dpiecewise,agadakos2019nibbler,ziegler2019honey}.
However, a recent study shows that these tools have many limitations~\cite{brown2024broad}.
They either fail to produce functionally correct libraries, require access to source code, or result in negligible—sometimes even negative—performance improvements.
From industry, the Rust programming language offers a feature-based dependency management system, where users explicitly declare required features, and only those are included at build time~\cite{rustlangFeaturesCargo}.
While effective, this strategy is not applicable to existing precompiled shared libraries.

In this paper, we argue that while discovering which parts of a shared library is used can be done using techniques such as dynamic analysis, it is challenging to reconstruct a working version of the shared library that reduces the size efficiently, and works robustly.
To address these limitations, we present \myapp, a tool that customizes shared libraries according to user-specific needs—without requiring access to source code or modifications to system components.
\myapp can be integrated with any tool for finding the used parts of a shared library, e.g.,  via dynamic tracing, runtime instrumentation for production workloads, or static analysis.
These identified code regions are specified as file ranges, which serve as input to \myapp.
Given this input, \myapp removes unused code, reorganizes the file layout, and produces a shared library fully compatible with existing systems.
To minimize file size and memory overhead, \myapp formulates the reorganization of file ranges as a variant of the classic \emph{Traveling Salesman Problem (TSP)} and applies a heuristic algorithm to find an approximately optimal layout.
It then updates program headers to ensure that only the required code regions are mapped into correct memory addresses at runtime.
Contrary to common belief that eliminating unused code yields negligible performance benefits—since such code is not executed~\cite{brown2024broad}, we show that \myapp delivers substantial improvements across \emph{all} four key metrics: reducing the shared library file size, memory usage, runtime overhead for cold starts, and enhancing security.
In summary, our contributions are as follows;
(1) We present \myapp, a novel tool for reconstructing shared libraries to retain user-required functionalities, without access to source code or modifications to system components; (2) We demonstrate that \myapp achieves significant improvements, reducing shared library size by up to \textbf{60\%}, lowering memory usage by up to \textbf{16\%}, and decreasing run time by up to \textbf{32\%}, all while preserving functional correctness and enhancing security; (3) We open-source \myapp.~\footnote{Tool will be open-sourced upon acceptance.}

\section{Background and Related Work}

\subsection{Shared Libraries: the ELF format}

Shared libraries in Linux use the Executable and Linkable Format (ELF)~\cite{oracleHeaderLinker}, which includes a list of program headers that define how the shared library are loaded into memory.
Basically, a program header can be viewed as a mapping from a file range (a contiguous sequence of bytes starting at a file offset) to a virtual memory address.
\begin{lstlisting}[language=json, label={lst:program-header}, caption={An example of program headers of a shared library. The file offsets and virtual memory addresses where the file offsets are loaded are the same. Some attributes of program headers are omitted for brevity.}]
Offset  VirtAddr FileSiz Align  ...
0x00000 0x00000  0x039e0 0x1000 ...
0x04000 0x04000  0x43519 0x1000 ...
0x48000 0x48000  0x398f0 0x1000 ...
...
\end{lstlisting}
Typically, the virtual address of a file range equals its file offset.
Listing\ref{lst:program-header} shows an example with three program headers:
the first maps the file range starting at 0x00000 (size 0x039e0) to virtual address 0x00000;
the second maps range 0x04000 (size 0x43519) to 0x04000; and so on.
Generally, file contents of a shared library can be moved to certain new locations, as long as the corresponding program headers are updated to map the file offsets to their original memory addresses~\cite{ziegler2019honey}.
\myapp also leverages this property to reconstruct shared libraries.

\subsection{Shared Library Debloating}
Many debloating tools have been proposed in the literature, mostly focusing on debloating executable binaries~\cite{navas2023occamv2,malecha2015occam,ahmad2021trimmer,sharif2018trimmer,brown2019carve,qian2019razor,koo2019configuration}.
Binary debloaters have a straightforward scope since each application typically involves a single binary.
These tools however typically ignore the shared libraries used by these binaries.
Generally, shared libraries can be much larger than the binaries that use them.
For example, the shared libraries used by the Nginx binary are 3$\times$ larger than the binary itself.
While there are a few tools that support shared library debloating, they typically require access to the source code to recompile the library~\cite{biswas2021ancile,quach2018dpiecewise},  fail to remove use case-specific bloat~\cite{porter2020blankit,agadakos2019nibbler}, or suffer from scalability issues, making them unsuitable for large shared libraries~\cite{ziegler2019honey}.
Our work focuses exclusively on the reconstruction of shared libraries and does not address bloat identification. 
We assume that users have already determined which parts of the library should be retained or removed—this identification can be performed using static or dynamic analysis techniques~\cite{bhattacharya2013combining,xu2014scalable,quach2019bloat}.
While \myapp can be integrated with software debloating tools to form a complete pipeline, it is outside the scope of this work.

\subsection{Program Loading and Cold Start}
In Linux, the system program loader (ld.so) is responsible for loading shared libraries into memory when a program is executed~\cite{man7Ldso8Linux}.
The program loader reads the program’s ELF headers to identify all dependent shared libraries and determine how each dependent shared library should be mapped into memory.
It proceeds to load each shared library by mapping the library into the memory address specified by the library's program headers by calling \texttt{mmap}.
This loading stage happens before the program execution begins.
As this library-loading occurs at program startup,
a program’s total \emph{run time} includes both \emph{loading time} and the \emph{execution time} of the program itself once all libraries are in place.
When an application depends on many or large shared libraries, this loading stage can impose significant overhead on the \emph{cold start} performance~\cite{tariq2025efficient}.
Cold starts are particular common in serverless and on-demand computing environments, where applications or functions are often launched from scratch for each new request or after a period of inactivity.
Therefore, optimizing the shared library loading process during cold starts is crucial for improving application responsiveness in such environments, as reducing library load time directly reduces the delay users experience before the application is ready to execute its logic~\cite{tariq2025efficient}.

\subsection{Problem Definition}\label{sec:problem-definition}
Consider a  shared library, denoted as $L_{in}$ that supports a wide range of features and a service that utilizes only one feature.
Through tracing the service in a production environment, the developer identifies the specific parts of $L_{in}$ that are executed to use the required feature
These are recorded as a list of non-overlapping ranges in the library, denoted by $R = \{r_1,r_2,\dots,r_n\}$, where $r_i.s$ and $r_i.e$ represent the start and end offsets of the $i$-th retained file range.
We assume the ranges in $R$ are sorted in ascending order by $r_i.s$.
The goal is to reconstruct a new shared library $L_{out}$ by retaining only the code in $R$ and removing all other unused code, while preserving the functional correctness of the  service.
More formally:

For simplicity in the following discussion, we adopt the following conventions:
(1) the list of file ranges $R$ is sorted by the starting offset $r_i.s$ and $r_i.s$ also equals the memory address where the file range $r_i$ is loaded;
(2) there is no overlap among file ranges in either the original or reconstructed shared libraries;
(3) we use curly brackets $\{\}$ to denote an ordered list; and
(4) we write $i \in \{\dots\}$ to indicate iteration over elements in an ordered list from the beginning to the end.

\section{Methodology}

In this section, we present \myapp, a tool and an algorithm that reconstructs shared libraries that preserves user-specified functionality while optimizing for file size, memory usage, and runtime performance. \myapp is based on an observation from studying various debloating tools; reconstructing the shared library from a set of file regions that have been identified as ``used'' has a tradeoff between memory usage and loading times. When debloating, each file range typically has a program header attached to it to map the file range to a virtual memory address. When there are many file regions accessed by a service, the increased number of program headers required for each region can increase the cold-start loading of the shared library, resulting in worse runtime performance for debloated libraries. Our tool aims to solve this problem by measuring the distance between two file ranges and defining a threshold for this distance $g$. If the distance between the ranges is smaller than $g$, then these ranges are stitched together, eliminating one of the program headers at the cost of extra memory usage.

\noindent \textbf{Overview of \myapp:}
\begin{algorithm}[hbt!]
    \SetAlgoLined
    \SetKwFunction{Merge}{Merge}
    \SetKwFunction{Shuffle}{Shuffle}

    \caption{\myapp}\label{alg:merge-shuffle}
    \Input{$L_{in}$: original shared library\\
        $R$: $\{r_1, \dots, r_n\}$, a list of file ranges\\
        $g$: a threshold parameter}
    \Output{$L_{out}$: reconstructed shared library}

    \BlankLine
    Replace all bytes in $L_{in}$ outside $R$ with \texttt{0xcc};

    $\Merge{$R$, $g$}$ \tcp*{(\S\ref{sec:merge})}

    $R' \leftarrow$ \Shuffle{$R$, $g$} \tcp*{(\S\ref{sec:shuffle})}

    \tcp{Create a new shared library}
    Create $L_{out}$ of size $\text{span}(R')$, filled with \texttt{0xcc};

    \ForEach{$r_i \in R$}{
        Copy bytes from $L_{in}$ at $r_i$ to $L_{out}$ at $r_i'$\;
        Create a program header mapping: $r_i' \rightarrow r_i$\;
    }
    \Return $L_{out}$\;
\end{algorithm}
Algorithm~\ref{alg:merge-shuffle} shows an overview of how \myapp works.
\myapp takes the shared library $L_{in}$, the file ranges $R$,  and the threshold parameter $g$ as input.
Line 1 replaces all bytes in $L_{in}$ that fall outside the specified ranges $R$ with the byte value \texttt{0xcc}, which corresponds to the INT3 instruction in the x86 architecture.
This \emph{cc-out} process mask out the unused code, reducing the attack surface.
Line 2 calls the \textsc{Merge} algorithm (see \S\ref{sec:merge}) is invoked to merge the ranges in $R$.
Specifically, any adjacent ranges in $R$ that have a gap not greater than the threshold $g$ are merged.
This step reduces the number of program headers required at lines 5-9, which create a program header for each file range in the reconstructed shared library.
Line 3 passes the merged ranges $R$ to the \textsc{Shuffle} algorithm (see \S\ref{sec:shuffle}), which shuffles them to reduce file size.
The output is a new set of ranges $R'$ with the new locations of $R$ in the reconstructed shared library $L_{out}$.
The two sets, $R$ and $R'$, are of equal length and correspond element-wise: the $i$-th file range in $R'$, denoted $r'_i$, is the new location of $r_i$, the $i$-th file range in $R$.

Line 4 creates the new shared library $L_{out}$, which is initialized to the size of the span of the shuffled ranges $R'$ and filled with \texttt{0xcc} bytes.
For each range $r_i$ in $R$, the corresponding bytes within the range are copied from $L_{in}$ to $L_{out}$ at the location $r'_i$ (Line 6).
A program header is then created to map $r'_i$ in the new file back to the original memory address $r_i$ (Line 7)\footnote{
As described in~\ref{sec:problem-definition}, each file range $r_i$ also represents the memory address to which that range is mapped when the shared library is loaded.}.
In doing so, the retained file ranges of the new shared library will be loaded into memory with the same memory addresses as the original shared library, ensuring that the functionality is preserved.
Also note that only the user-specified file ranges are mapped to their original memory addresses, while the rest of the file is not mapped, thus reducing the memory usage.

\begin{figure}
    \centering
    \includegraphics{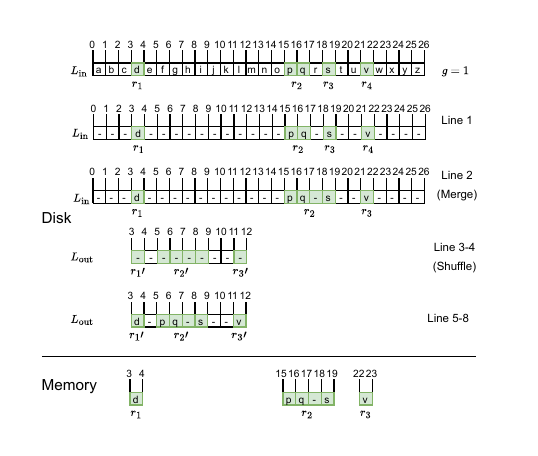}
    \caption{An example of the \textsc{MergeShuffle} algorithm. The input is a shared library of size 26 and four file ranges: $R = \{[3, 4), [15, 17), [18, 19), [21, 22)\}$, with a threshold $g = 1$.
    Line 1 performs cc-out by overwriting bytes outside $R$ with \texttt{0xcc} (shown as `-').
    Line 2 merges $r_2$ and $r_3$ as  the gap between them is not larger than $g$, resulting in $R = \{[3, 4), [15, 19), [21, 22)\}$.
    Line 3-4 shuffle the ranges to produce $R' = \{[3, 4), [5, 9), [11, 12)\}$ and creates the new shared library $L_{out}$ of size 9 bytes.
    Lines 5–8 copy data from $L_{in}$ at $r_i$ to $L_{out}$ at $r'_i$ and generate program headers that preserve original memory mappings.
    Only the user-specified file ranges are mapped to their original memory addresses, while the rest of the file is not mapped.
    }
    \label{fig:merge-shuffle-example}
\end{figure}
Figure~\ref{fig:merge-shuffle-example} illustrates an example of the \myapp process.
The \emph{cc-out} process ensures that all code outside the specified file ranges is removed, thereby reducing the attack surface.
The \mymerge step reduces the number of file ranges, which in turn reduce the number of program headers, thus reducing the loading overhead.
The \myshuffle step shuffles file ranges to reduce the overall file size.
By using program headers to map only the user-specified code regions to their original memory addresses, \myapp ensures that the reconstructed shared library preserves the original functionality while reducing memory usage.

\subsection{\mymerge: Balance Between Memory Usage and Run Time}\label{sec:merge}
As shown in Algorithm~\ref{alg:merge-shuffle}, each file range retained in the reconstructed shared library $L_{out}$ requires a corresponding program header to map it to its original memory address.
When the number of retained file ranges is large, this results in a large number of program headers, which can introduce non-negligible overhead when the shared library is loaded into memory.
According to this observation,  \mymerge merges two adjacent file ranges if they are separated by a small gap (not greater than a threshold $g$).
The rationale is that if two file ranges are close enough, it may be more efficient to merge them into a single range and use just one program header, even if it means mapping a few additional unused bytes into memory.
This merging operations reduce loading overhead caused by program headers, at the expense of slightly increased memory usage.

Algorithm~\ref{alg:merge} shows the pseudo-code of \mymerge, which is straightforward.
Apart from the user-specified file ranges, the algorithm takes as input a threshold parameter $g$.
Any two adjacent file ranges whose gap is less than or equal to $g$ will be merged into one continuous range.
This reduces the total number of ranges—and hence program headers—required in the final library.
\begin{algorithm}[hbt!]
    \SetAlgoLined

    \caption{Merge}\label{alg:merge}
    \Input{$R$: $\{r_1, \dots, r_n\}$, a list of file ranges\\
        $g$: a threshold parameter}
    \For{$i=2; i\leq{n}; i+=1$}{
        \If{$r_i.s - r_{i-1}.e \leq g$}{
            $r_{i-1}.e = r_i.e$\;
            Remove $r_i$ from $R$\;
        }
    }
\end{algorithm}
The  merging operation introduces a trade-off between memory usage and run time, controlled by the parameter $g$.
When $g = 0$, no merging occurs; each file range is mapped individually, resulting in minimal memory usage but also the highest number of program headers and thus the highest potential loading time caused by these headers.
At the other extreme, if $g$ is large enough to merge all ranges into one, the resulting memory usage is equivalent to the original library, but the number of program headers is minimized to the same as the original shared library.
However, increasing $g$ does not always lead to faster loading.
While fewer program headers reduce loading overhead, the increased bytes to be loaded may offset these gains.
This trade-off between loading time and memory usage is a critical design consideration in \mymerge.
The threshold parameter $g$ enables users to control the trade-off between memory usage and run time, allowing them to tailor the reconstruction based on whether their application is more sensitive to memory usage or runtime efficiency.

\subsection{\myshuffle: Reduce File Size under Congruence Constraint}\label{sec:shuffle}
In Algorithm~\ref{alg:merge-shuffle}, the \myshuffle algorithm takes as input the merged file ranges $R = \{r_1, \dots, r_n\}$ produced by the \mymerge step, along with a threshold parameter $g$.
We begin by formulating the shuffle problem from a theoretical perspective, and then introduce two algorithms to solve it.
We then discuss practical considerations that arise when applying these algorithms, where the value of $g$ limits the usage of the two algorithms.
Finally, we present the complete \myshuffle algorithm used in \myapp, which apply different shuffle algorithms based on the threshold parameter $g$.

\noindent\textbf{The Range Shuffling Problem:}
The goal of the \myshuffle algorithm is to shuffle file ranges in $R$ to minimize the final file size of the reconstructed shared library.
Specifically, for each file range $r_i \in R$, we need to move it to a new location, denoted as $r'_i$, such that the span of the new file ranges $R' = \{r'_1, \dots, r'_n\}$ is minimized.
Each new file ranges $r'$ should satisfy the congruence constraint, i.e., the start offset of $r'$ should have the same modulo to the system page size $P$ as the original start offset of $r$~\footnote{This is according to the ELF specification. We explain the constraint in more details in the appendix},
\begin{equation}
    \label{eq:congruence-shuffled}
    r'_i.s \equiv r_i.s \pmod{P}
\end{equation}
We refer to the above problem of shuffling file ranges under the congruence constraint to minimize the span of the new file ranges as the Range Shuffling Problem (RSP).
We next present two algorithms to solve the RSP approximately: \linearshuffle and \tspshuffle.

\subsubsection{\linearshuffle}
\begin{figure}
    \centering
    \includegraphics{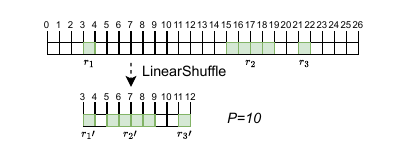}
    \caption{An example of the \textsc{LinearShuffle} algorithm. The input is a list of file ranges: $R = \{[3, 4), [15, 19), [21, 22)\}$.
    The algorithm begins by keeping the first range $[3, 4)$ unchanged.
    Since the gap between the end of the first range (4) and the start of the second range (15) is greater than the page size, the second range is shifted forward by $\lfloor \frac{15 - 4}{10} \rfloor \times 10 = 10$ bytes, resulting in the new range $[5, 9)$.
    This movement increases the gap between the second and third ranges to 12 bytes.
    As this new gap also exceeds the page size, the third range is similarly moved forward by $\lfloor \frac{21 - 9}{10} \rfloor \times 10 = 10$ bytes, yielding the new range $[11, 12)$.
    As a result of these shifts, the total file size is reduced from 26 bytes to 9 bytes, while preserving the required congruence constraints.
    }
    \label{fig:linear-shuffle-example}
\end{figure}

The \linearshuffle algorithm is a greedy, linear-time complexity strategy.
The pseudocode is shown in Algorithm~\ref{alg:linear-shuffle}.
A file range $r_i$ will be moved forward by $\lfloor{\frac{r_{i}.s-r_{i-1}.e}{P}}\rfloor\times {P}$ bytes.
The reason is that since under the congruence constraint $r'_{i}.s \equiv r_i.s \pmod{P}$, we have $r'_{i}=r_i-k*P$, where $k$ is a non-negative integer.
In other words, a file range will not be moved if the gap between two adjacent file ranges is smaller than the page size ($k=0$).
A file range will be moved forward by an integer multiple of the page size, if the gap is larger than the page size $k>0$;
By moving the file range forward, it is closer to the previous file range and the total size of the shared library is reduced.
Also note that after the file range is moved forward, the space that the file range occupied before will also be zeroed out and treated as a new gap (lines 8-9).
In doing so, the congruence constraint is met, and the size of shared library is reduced.
Figure~\ref{fig:linear-shuffle-example} illustrates an example of the \linearshuffle algorithm.

\begin{algorithm}[hbt!]
    \SetAlgoLined
    \caption{\linearshuffle}\label{alg:linear-shuffle}
    \Input{ $R = \{r_1,\dots,r_n\}$}
    \Output{$R' = \{r'_1,\dots,r'_n\}$}
    $r'_1=r_1$\;
    $R'=\{r'_1\}$, $e=r_1.e$\;
    \For{$i=2;i<=n;i+=1$}{
    $r'_i.s  = r_i.s - \lfloor{\frac{r_i.s-r_{i-1}.e}{P}}\rfloor\times {P}$\;
    $r'_i.e = r'_i.s + r_i.e - r_i.s$\;
    $R' = R' \cup \{[r'_i)\}$\;
    $e=r'_i.e$\;
    }
    \Return $R'$\;
\end{algorithm}

\subsubsection{\tspshuffle}
We show that RSP is equivalent to the classic Traveling Salesman Problem (TSP), which is NP-hard, and thus can be  addressed using heuristic or approximate TSP-solving algorithms~\cite{junger1995traveling}.
To minimize the final span of the shuffled file ranges, and given that the sizes of the ranges are fixed, the problem equals to minimizing the gaps between file ranges.
We model this using a directed complete graph $G=(V,E)$, where each vertex $v_i$ is a file range $r_i$ in $R$.
For each pair of distinct vertices $(v_i, v_j)$, we define a directed edge $e_{ij} \in E$ with weight $w_{ij}$ representing the cost (i.e., the gap) of placing $r_j$ \emph{immediately} after $r_i$ under the congruence constraint.
Therefore, the edge weight is defined as:
\begin{equation}
    w_{ij} = (r_j.s - r_i.e) \% P
\end{equation}
Minimize the total file size is equivalent to finding a Hamiltonian path in this directed graph that visits each vertex exactly once and minimizes the total weight of the edges in the path.
RSP is thus reducible to TSP.
This path determines the optimal shuffling of file ranges that minimizes total span while satisfying the congruence constraints.

\begin{figure}
    \centering
    \includegraphics{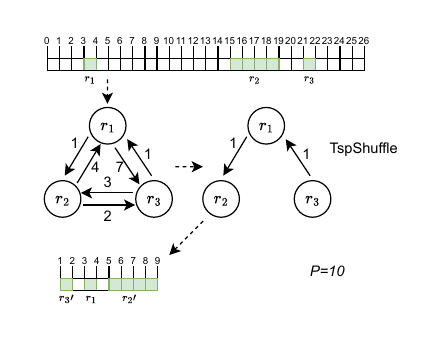}
    \caption{An example of the \textsc{TspShuffle} algorithm. The input is a list of file ranges: $R = \{[3, 4), [15, 19), [21, 22)\}$.
    The page size $P$ is set to 10 for this example.
    The algorithm first creates a complete directed graph with three vertices, each representing a file range.
    And the weight of each edge is calculated by line 5 of Algorithm~\ref{alg:tsp-shuffle}.
    Then the threshold-accepting algorithm is used to find a Hamiltonian path that visits each vertex exactly once and minimizes the total weight of the edges in the path.
    The resulting path is $H = \{r_3, r_1, r_2\}$, which specifies the order in which the file ranges should be placed.
    Starting from offset 0, $r_3$ is placed first at the next immediately available location that satisfies the congruence constraint with $r_3.s$; in this example, that location is offset 1.
    Next, $r_1$ is placed at the next immediately available offset congruent to $r_1.s$, which is offset 3.
    Finally, $r_2$ is placed at the next available offset congruent to $r_2.s$, which is offset 5.
    The file size is reduced from 26 bytes to 8 bytes, smaller than the \linearshuffle algorithm.
    }
    \label{fig:tsp-shuffle-example}
\end{figure}

Algorithm~\ref{alg:tsp-shuffle} presents the pseudocode for the \tspshuffle algorithm.
The process begins by constructing a complete directed graph $G = (V, E, w)$, as we described above (Lines 1–6).
Next, the algorithm calls a threshold-accepting algorithm to find a path $H$ through the graph (Line 7).
Threshold-accepting is a metaheuristic algorithm for solving combinatorial optimization problems, particularly TSP~\cite{dueck1990threshold}.
This path visits each vertex, i.e., each file range, exactly once and approximates the minimum total edge weight.
Lines 8–14 use the path $H$ to compute the new file ranges $R'$.
For each range $r_i$ in the order specified by $H$, starting from zero, the algorithm places it at the first available file offset that satisfies the congruence constraint and does not overlap with previously placed ranges.
Figure~\ref{fig:tsp-shuffle-example} shows an example of \tspshuffle.

\begin{algorithm}[hbt!]
    \SetAlgoLined
    \caption{\tspshuffle}\label{alg:tsp-shuffle}
    \Input{ $R = \{r_1,\dots,r_n\}$}
    \Output{$R' = \{r'_1,\dots,r'_n\}$}
    \Comment{Create the graph}
    $G=(V,E, w)$\, where $V$ is the set of vertices, $E$ is the set of edges, both initialized to empty sets; $w$ is the weight function that maps each edge to its weight\;
    \For{$i=1;i<=n;i+=1$}{
        $V=V \cup \{r_i\}$\;
        $E=E \cup \{(r_i,r_j) | j\in {1,2,\dots,n}, j\neq i\}$\;
        $w(r_i,r_j) = (r_j.s - r_i.e) \mod P$\;
    }
    \Comment{Find the Hamiltonian path}
    $H= threshold\_accepting(G)$\;
    \Comment{Generate the new file ranges}
    $R'=\{\}$,  $e=0$\;
    \For{$r_i \in H$}{
        $r'_i.s=e+(r_i.s-e)\mod P$\;
        $r'_i.e=r'_i.s+r_i.e-r_i.s$\;
        $R'=R'\cup {r'_i}$\;
        $e=r'_i.e$\;
    }
    \Return $R'$\;
\end{algorithm}

\subsubsection{\myshuffle}
Although both \linearshuffle and \tspshuffle are effective in theory, the way operating systems load shared libraries into memory introduces practical constraints that affect their applicability.
Specifically, when the system maps a file range into memory, it aligns the start offset of the range down to the nearest page boundary.
This behavior introduces a risk of range overlap in memory when the $g$ is smaller than the system page size $P$ for \tspshuffle, as it may change the relative order of file ranges.
This can lead to corrupted memory, where content from one range is unintentionally overwritten by another, resulting in incorrect program behavior.
\linearshuffle preserves the relative order of the file ranges, thus avoiding this problem~\footnote{We provide a detailed explanation of the range overlap problem and why \linearshuffle avoids it in the appendix.}.

To address this, the final \myshuffle algorithm adapts to the value of $g$. When $g < P$, \myapp uses \linearshuffle to prevent page overlap.
When $g \geq P$, both algorithms are safe to use, but we select \tspshuffle because it typically achieves better file size reduction.
The final \myshuffle logic used in Algorithm~\ref{alg:merge-shuffle} is shown in Algorithm~\ref{alg:shuffle}.

\begin{algorithm}[hbt!]
    \SetAlgoLined
    \caption{\myshuffle}\label{alg:shuffle}
    \Input{$R = \{r_1, \dots, r_n\}$, $g$}
    \Output{$R' = \{r'_1, \dots, r'_n\}$}
    \If{$g < P$}{
        $R' = \text{\linearshuffle}(R)$\;
    }\Else{
        $R' = \text{\tspshuffle}(R)$\;
    }
    \Return $R'$\;

\end{algorithm}

\noindent\textbf{Implementation:} We implement \myapp in two components: the algorithm component and the rewriting component.
The algorithm component is implemented in approximately 400 lines of Python.
It takes as input a list of file ranges and a gap threshold, and produces a reordered list of file ranges.
The rewriting component is implemented in approximately 4500 lines of Rust.
It takes the original shared library and the list of file ranges output by the algorithm componenet as input, and outputs a reconstructed shared library containing only the specified file ranges.

\section{Experiments}

To evaluate the effectiveness of \myapp, we conduct experiments on applications from diverse domains that make use of shared libraries.
The selected applications are listed in Table~\ref{tab:apps}.
Our evaluation includes:
A machine learning framework, Torch~\cite{paszke2019pytorch}, evaluated under two distinct use cases: inference and training.
Three command-line tools drawn from the Debloater-Eval-Benchmark~\cite{brown2024broad}, where we use the representative use cases as defined in the original benchmark.
Four microservice applications selected from DeathStarBench~\cite{gan2019open}, a widely used microservice benchmark suite, again using the original benchmark’s use cases.
To collect shared library usage information for each application, we use RTrace~\cite{rtrace}.
We execute each application with selected representative use cases while tracing with RTrace, which outputs the file ranges of shared libraries actually accessed during execution.
Our evaluation focuses on four key metrics: file size, memory usage, run time, and security.
For run time, we focus solely on the \emph{cold start} — when the binary and shared libraries are first loaded from disk into memory.

\begin{table}[t]
    \centering
    \caption{Evaluated Applications and Use Cases}
    \label{tab:apps}
    \begin{tabular}{ll}
        \toprule
        \textbf{Application} & \textbf{Use Case}                                   \\
        \midrule
        \multicolumn{2}{l}{\textit{Machine Learning}}                              \\
        Torch-Inference      & CNN model inference                                 \\
        Torch-Train          & CNN model training                                  \\

        \midrule
        \multicolumn{2}{l}{\textit{Command-Line Tools (Debloater-Eval-Benchmark)}} \\
        ImageMagick          & Image processing (convert, resize, crop, etc.)      \\
        Nmap                 & Network scanning (TCP, service, report, etc.)       \\
        Poppler              & PDF conversion (simple, complex, page ranges)       \\

        \midrule
        \multicolumn{2}{l}{\textit{Microservices (DeathStarBench)}}                \\
        ComposePostService   & Post send/receive                                   \\
        HomeTimelineService  & Home timeline read/write                            \\
        TextService          & Text send/receive                                   \\
        UserService          & User registration/login                             \\
        \bottomrule
    \end{tabular}
\end{table}

\subsection{Overview of Size, Memory and Run time Reduction}\label{sec:res-overview}
We first run each application with its original shared libraries and measure three metrics: the total size of shared libraries, peak memory usage, and run time.
We then run the same applications with the reconstructed shared libraries and collect the same metrics.
The relative values are computed as the ratio of the reconstructed version to the original; values below 100\% indicate improvements, while values above 100\% indicate deterioration in the respective metric.
Both original run and reconstructed run are executed 20 times to obtain mean and standard deviation values.

Table~\ref{tab:overview-results} presents an overview of the results under the most aggressive reconstruction setting, where the threshold $g$ is set to 0.
This configuration maximizes code removal and memory savings but may increase run time overhead.
The number of reconstructed shared libraries varies across applications.
In general, the reconstructed libraries are significantly smaller, with relative sizes ranging from 40.3\% to 65.8\%.
This size reduction leads to lower memory usage, with relative peak memory ranging from 83.7\% to 98.6\%.
Runtime performance also improves in most cases, with relative run times ranging from 67.6\% to 100.1\%.

While all applications show significant size reductions, memory and runtime improvements vary.
This is expected, as the benefits of code removal depend on each application's characteristics.
In data-intensive workloads, most memory is occupied by data, making code reduction less impactful on memory usage.
Similarly, for applications taking a long time to run, library loading time is negligible compared to total execution time, making run time improvements less significant.
Also due to this reason, more (less) memory reduction does not always translate to more (less) run time improvement.
For instance, Imagemagick shows only a 1.4\% memory reduction but a 10.3\% run time improvement, while Poppler achieves 12\% memory reduction with just a 2.8\% run time gain.

\begin{table}[h]
    \centering
    \caption{Overview of number of libraries, relative size, memory and run time of applications running using original and reconstructed shared libraries.
        Values are relative to the original version, averaged over 20 runs, with standard deviation shown.
        $g=0$.}
    \label{tab:overview-results}
    \scalebox{0.93}{
        \begin{tabular}{lrrrr}
            \toprule
            Application         & \#Lib. & Size(\%) & Peak Memory(\%)           & Run Time(\%)     \\
            \midrule
            Torch-Inference     & 7      & 40.3     & 83.7           $\pm$ 0.08 & 67.6  $\pm$ 2.07 \\
            Torch-Train         & 7      & 40.6     & 92.3           $\pm$ 0.02 & 98.6  $\pm$ 1.23 \\
            Imagemagick         & 8      & 59.6     & 98.6           $\pm$ 0.03 & 89.7  $\pm$ 1.56 \\
            Nmap                & 5      & 61.3     & 96.5           $\pm$ 1.21 & 100.1 $\pm$ 0.82 \\
            Poppler             & 12     & 65.8     & 88.0           $\pm$ 0.71 & 97.2  $\pm$ 2.31 \\
            ComposePostService  & 9      & 62.8     & 84.4           $\pm$ 0.82 & 90.1  $\pm$ 6.02 \\
            HomeTimelineService & 9      & 62.8     & 84.8           $\pm$ 0.68 & 92.2  $\pm$ 4.13 \\
            TextService         & 9      & 62.8     & 84.3           $\pm$ 0.76 & 88.7  $\pm$ 5.37 \\
            UserService         & 24     & 64.1     & 81.6           $\pm$ 0.51 & 93.7  $\pm$ 3.31 \\
            \bottomrule
        \end{tabular}
    }
\end{table}

\subsection{Impact of Threshold $g$}\label{sec:vary-g}

From the results above, setting $g = 0$ consistently leads to substantial reductions in file size.
However, improvements in memory usage and run time vary across applications.
In this section, we analyze how tuning the threshold $g$ affects all three metrics—file size, memory usage, and run time.

We begin by computing the gap sizes between adjacent file ranges in the input file ranges $R$.
We then select various percentiles of these gap sizes as candidate values for $g$.
Specifically, we use the 0th, 10th, 20th, ..., 90th percentiles.
Additionally, we sample more densely between the 90th and 100th percentiles, including 91st, 92nd, ..., 99th percentiles.
This aims to gain finer-grained insight in the region where the three metrics varies more pronounced.
We exclude the 100th percentile, as it effectively corresponds to using the original shared libraries without any reconstruction.
For each value of $g$, we run the original and reconstructed applications 20 times and report the mean and standard deviation of the results.
The absolute values of $g$ and corresponding percentiles are shown in Table~\ref{tab:g-values} in the appendix.

Figure~\ref{fig:gap-change} illustrates how relative file size, peak memory usage, and run time vary with different values of $g$.
For file size and memory, increases in $g$ when it is small have little impact, but larger values of $g$ result in noticeable increases across all applications.
Run time is more complex.
Increasing $g$ reduces the number of program headers, which may reduce loading overhead.
However, it also decreases the memory savings, which can negatively impact performance.
These competing effects mean that adjusting $g$ can either improve or degrade run time, as shown in the figure.
Notably, some applications exhibit threshold values that offer a favorable trade-off—retaining most of the size and memory savings while further improving run time.
For instance, in Figure~\ref{fig:home-timeline-service}, increasing $g$ from the 10th to the 80th percentile reduces relative run time from 93\% to 85\%, with minimal impact on file size and memory.
This suggests that $g$ can be effectively tuned to optimize performance for specific applications.

\begin{figure*}[htbp]
    \centering
    \begin{subfigure}[b]{0.325\textwidth}
        \includegraphics[width=\textwidth]{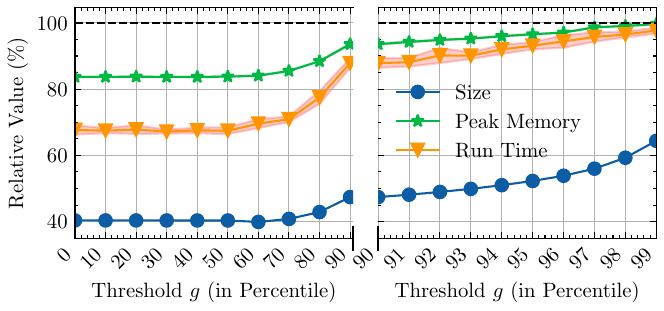}
        \caption{Torch-Inference}
        \label{fig:inference}
    \end{subfigure}
    \hfil
    \begin{subfigure}[b]{0.325\textwidth}
        \includegraphics[width=\textwidth]{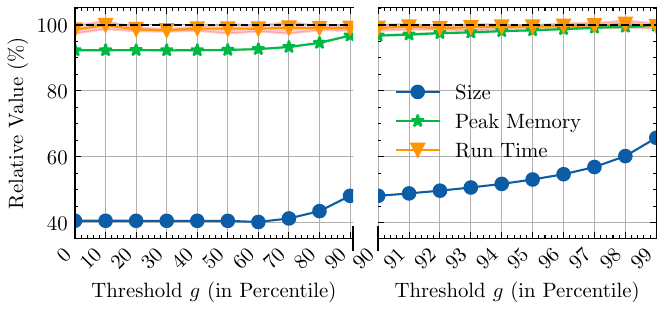 }
        \caption{Torch-Train}
        \label{fig:train}
    \end{subfigure}
    \hfil
    \begin{subfigure}[b]{0.325\textwidth}
        \includegraphics[width=\textwidth]{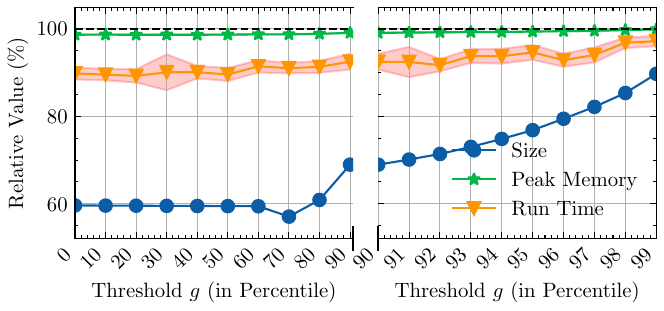}
        \caption{imagemagick}
        \label{fig:imagemagick}
    \end{subfigure}
    \vskip\baselineskip
    \begin{subfigure}[b]{0.325\textwidth}
        \includegraphics[width=\textwidth]{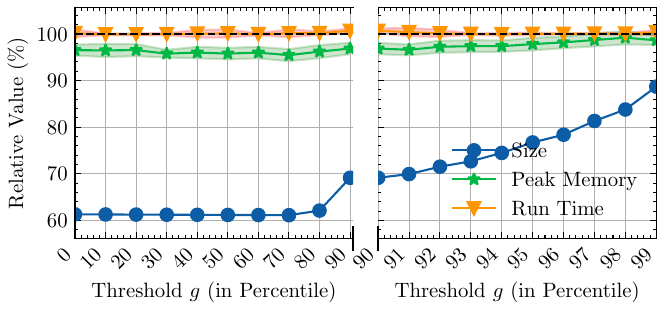}
        \caption{nmap}
        \label{fig:nmap}
    \end{subfigure}
    \hfil
    \begin{subfigure}[b]{0.325\textwidth}
        \includegraphics[width=\textwidth]{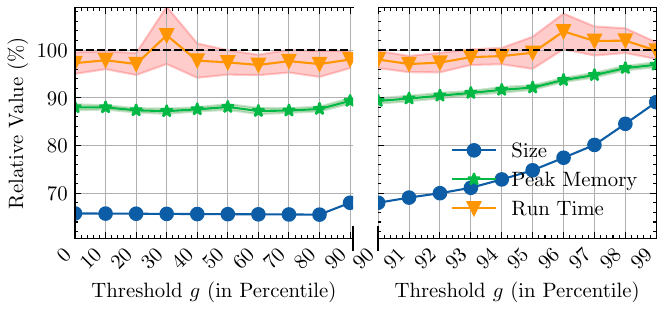}
        \caption{poppler}
        \label{fig:poppler-compression}
    \end{subfigure}
    \hfil
    \begin{subfigure}[b]{0.325\textwidth}
        \includegraphics[width=\textwidth]{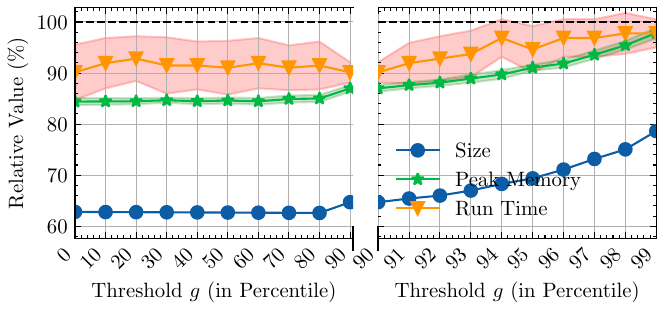}
        \caption{Compose Post Service}
        \label{fig:compose-post-service}
    \end{subfigure}
    \vskip\baselineskip
    \begin{subfigure}[b]{0.325\textwidth}
        \includegraphics[width=\textwidth]{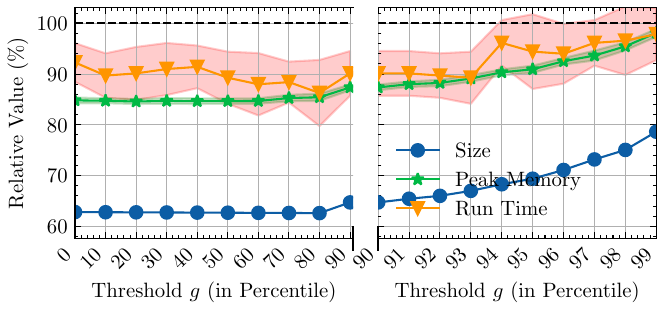}
        \caption{Home Timeline Service}
        \label{fig:home-timeline-service}
    \end{subfigure}
    \hfil
    \begin{subfigure}[b]{0.325\textwidth}
        \includegraphics[width=\textwidth]{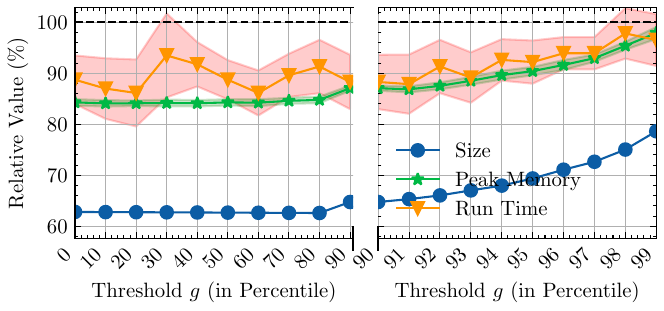}
        \caption{Text Service}
        \label{fig:text-service}
    \end{subfigure}
    \hfil
    \begin{subfigure}[b]{0.325\textwidth}
        \includegraphics[width=\textwidth]{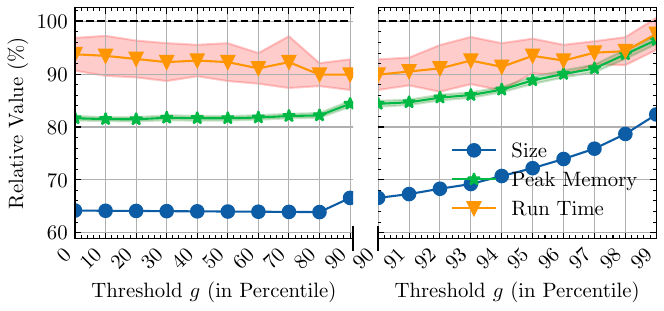}
        \caption{User Service}
        \label{fig:user-timeline-service}
    \end{subfigure}
    \caption{The relative values of file size, memory usage and run time of applications running using original and reconstructed shared libraries with different $g$.}
    \label{fig:gap-change}
    \vspace{-1em}
\end{figure*}

\subsection{Analysis of Size Reduction}\label{sec:exp-size}
This section analyzes the file size reduction achieved by the reconstructed shared libraries.
Since \myapp only removes unused code, we focus on the code size of the shared libraries.
Specifically, we measure how much code is removed from the original libraries using the \textsc{LinearShuffle} and \textsc{TspShuffle} algorithms.
To ensure a fair comparison, we fix the threshold $g$ to 4096 bytes, under which both algorithms are able to produce functional shared libraries.
We also compute the maximal size reduction, defined as the theoretical maximum reduction achievable by removing all unused code without enforcing the congruence constraint.
This serves as an upper bound for evaluating how closely the two algorithms approach the optimal result.

Table~\ref{tab:code-size} presents the original code size, the maximal size reduction along with the actual size reduction achieved by both \linearshuffle and \tspshuffle for each application.
Overall, both shuffle algorithms achieve reductions close to the maximal size reduction, over 70\% reduction for all applications.
Moreover, \tspshuffle consistently outperforms \linearshuffle, demonstrating its advantage in producing more compact library layouts.
This code reduction directly contribute to the memory and run time improvements as we will discuss in the next sections.

\begin{table}[h]
    \centering
    \caption{Comparison of code size reductions achieved by \linearshuffle and \tspshuffle. `
        `Ori." denotes the original code size; ``Max." represents the theoretical upper bound on size reduction; values in parentheses show the reduction in percentage. $g=4096$.}
    \label{tab:code-size}
    \scalebox{0.95}{
        \begin{tabular}{lrrrrr}
            \toprule
            Application         & \#Lib. & Ori. & Max./MB    & Linear/MB        & TSP/MB    \\
            \midrule
            Torch-Inference     & 7      & 273  & 262   (96) & 256         (94) & 260  (95) \\
            Torch-Train         & 7      & 273  & 261   (96) & 255         (93) & 259  (95) \\
            Imagemagick         & 8      & 4    & 4     (89) & 3           (78) & 4    (82) \\
            Poppler             & 12     & 4    & 3     (90) & 3           (82) & 3    (83) \\
            Nmap                & 5      & 3    & 3     (97) & 3           (94) & 3    (95) \\
            ComposePostService  & 9      & 4    & 3     (89) & 3           (79) & 3    (81) \\
            HomeTimelineService & 9      & 4    & 3     (90) & 3           (79) & 3    (81) \\
            TextService         & 9      & 4    & 3     (90) & 3           (79) & 3    (81) \\
            UserService         & 24     & 6    & 5     (82) & 4           (70) & 4    (72) \\
            \bottomrule
        \end{tabular}
    }
\end{table}

\subsection{Analysis of Memory Reduction}\label{sec:memory}
As shown in the previous section, \myapp achieves substantial code size reduction.
This reduction not only decreases the on-disk size of shared libraries but also significantly impacts memory usage, as less code is mapped into memory at runtime.
In this section, we analyze the memory reduction effect using the Torch-Inference application with the threshold of $g = 0$, as this application achieved the highest file size reduction among all evaluated applications as shown in Table~\ref{tab:overview-results}.

\begin{figure}[ht]
    \begin{minipage}[t]{0.24\textwidth}
        \centering
        \includegraphics[width=0.95\textwidth]{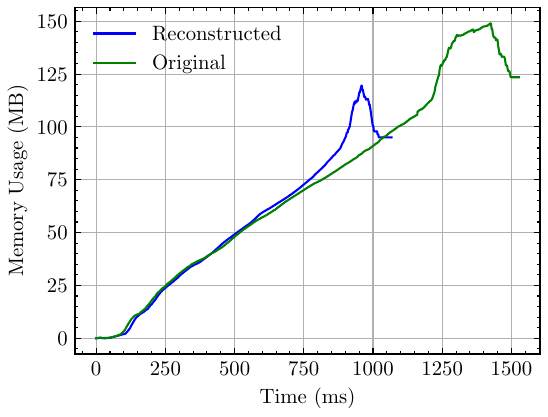}
        \caption{Memory footprint of Torch-Inference running using original and reconstructed shared libraries.
            The memory values at each time point are averaged over 20 runs.
            $g=0$.}
        \label{fig:mem-footprint}
    \end{minipage}
    \hfill
    \begin{minipage}[t]{0.24\textwidth}
        \includegraphics[width=\textwidth]{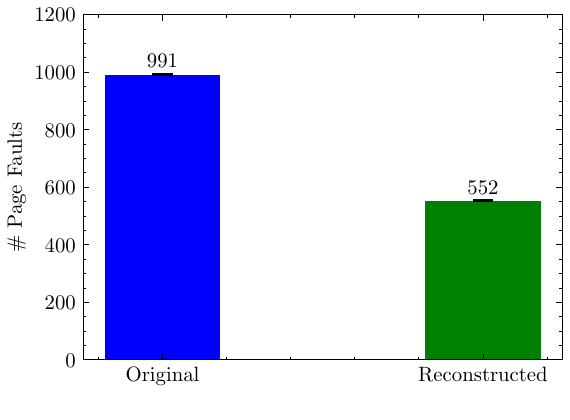}
        \caption{Number of page fault during the execution of Torch-Inference using the original and reconstructed shared libraries.
            The values are averaged over 20 runs.
            $g=0$.}
        \label{fig:page-fault}
    \end{minipage}
\end{figure}

\textbf{Memory Reduction.}
Figure~\ref{fig:mem-footprint} shows the memory footprint of Torch-Inference when executed with the original and reconstructed shared libraries.
The results show a clear reduction in memory usage: peak memory decreases from approximately 150 MB to 125 MB—a 17\% improvement.
Table~\ref{tab:mem-analysis} provides detailed memory usage of the original and reconstructed shared libraries.
Reconstructed libraries occupy significantly fewer physical pages in memory.
For example, the ibtorch\_cpu.so library, which is the largest shared library in the Torch-Inference application, reduces from 15,648 physical pages to 7,020 pages when reconstructed (a 55\% reduction).
This directly translates to reduced memory usage.

\textbf{Memory Locality.}
Beyond reducing memory consumption, the reconstructed shared libraries also exhibit improved locality.
As shown in Table~\ref{tab:mem-analysis}, the original libraries have a VMA-per-page ratio equal to 1, indicating that each physical page only contains a single virtual memory area (VMA).
In contrast, the reconstructed libraries achieve a VMA-per-page ratio of more than 1 for some libraries.
For example, the libtorch\_cpu.so library has a VMA-per-page ratio of 1.07, meaning that some pages contain multiple VMAs.
We observed that some pages even contain up to five VMAs.
This means reconstructed shared libraries have better \emph{memory locality}, which can improve run time performance.
Because when multiple frequently accessed code regions are co-located within the same physical page,  memory caches are more likely to capture relevant instructions together.
As a result, accessing one function may implicitly load others into cache, reducing cache misses and improving execution efficiency.
This effect is supported by the observed reduction in page faults during execution as shown in Figure~\ref{fig:page-fault}.
Using the reconstructed shared libraries significantly decreases the number of page faults—from 991 to 552—a reduction of approximately 44\%.
This reduction in page faults contributes directly to the observed run time improvements, as we will discuss in the next section.

\begin{table}[h]
    \centering
    \caption{Memory information of the original and reconstructed shared libraries. $g=0$.}
    \label{tab:mem-analysis}
    \begin{tabular}{lrrrr}
        \toprule
                      & \multicolumn{2}{c}{\#Physical pages} & \multicolumn{2}{c}{VMA-per-page}                            \\
        \cmidrule(l{6pt}r{6pt}){2-3} \cmidrule(l{6pt}r{6pt}){4-5}
        Library       & Original                             & Reconstructed                    & Original & Reconstructed \\
        \midrule
        libtorch\_cpu & 15648                                & 7020                             & 1.00     & 1.07          \\
        libstdc++     & 368                                  & 256                              & 1.00     & 1.06          \\
        libc10        & 236                                  & 149                              & 1.00     & 1.11          \\
        libgomp       & 43                                   & 14                               & 1.00     & 1.00          \\
        libgcc\_s     & 25                                   & 10                               & 1.00     & 1.00          \\
        librt-2.31    & 8                                    & 5                                & 1.00     & 1.00          \\
        libtorch      & 1                                    & 1                                & 1.00     & 1.00          \\
        \bottomrule
    \end{tabular}
\end{table}

\subsection{Analysis of Run Time Reduction}\label{sec:runtime}
As discussed in the previous section, memory reduction can influence the run time performance of applications.
In this section, we further investigate this effect by analyzing the runtime of applications using both the original and reconstructed shared libraries.
We again use the Torch-Inference application as a representative example.
The run time of a program consists of two components:  loading time and  execution time.
Loading time refers to the duration required to load the binary and its dependent shared libraries into memory before invoking the program's main function.
Execution time refers to the time spent running the main function.
We utilize LD\_DEBUG~\cite{oracleDebuggingLibrary} to measure loading time, and we modify the main function in Torch-Inference to log the execution time.

\begin{figure}[t]
    \centering
    \begin{subfigure}[b]{0.15\textwidth}
        \centering
        \includegraphics[width=\textwidth]{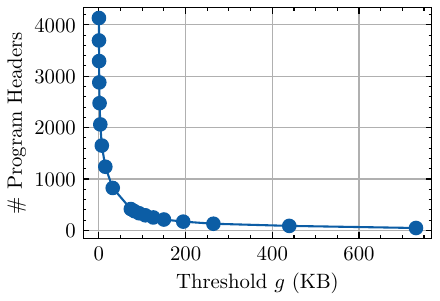}
        \caption{\#Program Header}
        \label{fig:num-phdr}
    \end{subfigure}
    \begin{subfigure}[b]{0.15\textwidth}
        \centering
        \includegraphics[width=\textwidth]{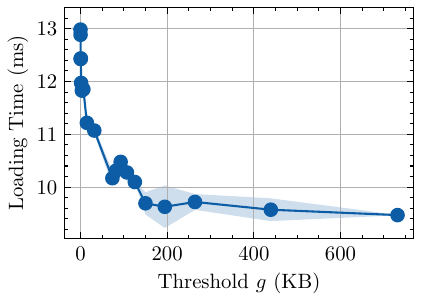}
        \caption{Loading Time}
        \label{fig:load-time}
    \end{subfigure}
    \begin{subfigure}[b]{0.15\textwidth}
        \centering
        \includegraphics[width=\textwidth]{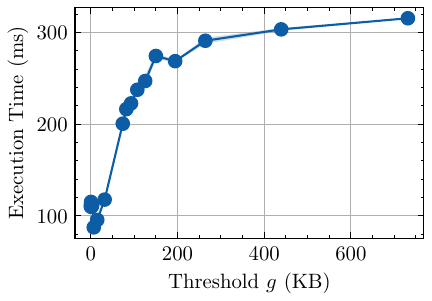}
        \caption{Execution Time}
        \label{fig:execution-time}
    \end{subfigure}
    \caption{Loading time and execution time of Torch-Inference running using original and reconstructed shared libraries with different $g$.}
    \label{fig:run-time-analysis}
\end{figure}

Figure~\ref{fig:run-time-analysis} shows how the number of program headers, loading time and execution time vary with different threshold $g$.
The values of $g$ are sampled from the 0th to the 99th percentiles of the gap sizes, as described in \S\ref{sec:vary-g}.
The number of program headers decreases as $g$ increases, which is expected since larger values of $g$ allow more file ranges  to be merged together.
The decrease in program headers leads to a reduction in loading time, as fewer headers need to be processed during the loading phase.
However, execution time tends to increase, since higher values of $g$ reduce memory savings and lead to more code being loaded into memory and worse locality.
When $g$ becomes sufficiently large, the reconstructed shared library is equal to the original one.
Execution time improvements are most noticeable during the first execution of code.
In iterative workloads like Torch-Train, the performance gain is primarily observed in the initial iteration, while later iterations see minimal impact due to memory caching.

\subsection{Reconstructed Yet Shareable}\label{sec:exp-shared}
A key advantage of shared libraries is their ability to be used by multiple applications.
In previous sections, each shared library was reconstructed specifically for a single application, resulting in multiple specialized versions that are no longer shareable.
In this section, we demonstrate that \myapp can also produce a shared version of a reconstructed library that remains shareable across multiple applications.

To achieve this, \myapp reconstructs a shared library using the union of the input file ranges collected from multiple applications.
We evaluate this using the four microservices, which share many common libraries.
While this approach retains more code, it enables the resulting shared library to be shared by all related applications.

\begin{figure}[t]
    \centering
    \begin{subfigure}[b]{0.15\textwidth}
        \centering
        \includegraphics[width=\textwidth]{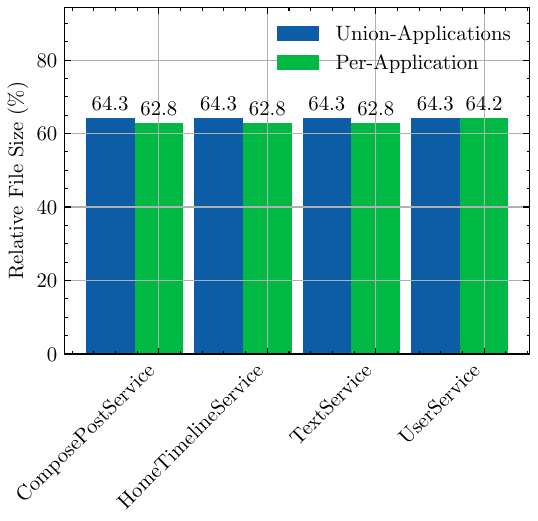}
        \caption{Relative Size}
        \label{fig:all-separate-size}
    \end{subfigure}
    \begin{subfigure}[b]{0.15\textwidth}
        \centering
        \includegraphics[width=\textwidth]{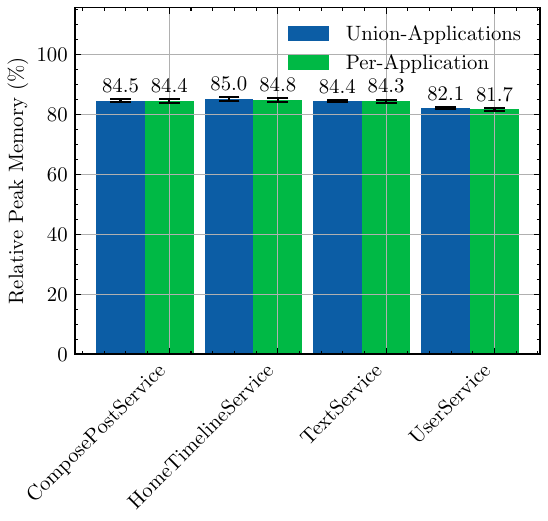}
        \caption{Relative Memory}
        \label{fig:all-separate-mem}
    \end{subfigure}
    \begin{subfigure}[b]{0.15\textwidth}
        \centering
        \includegraphics[width=\textwidth]{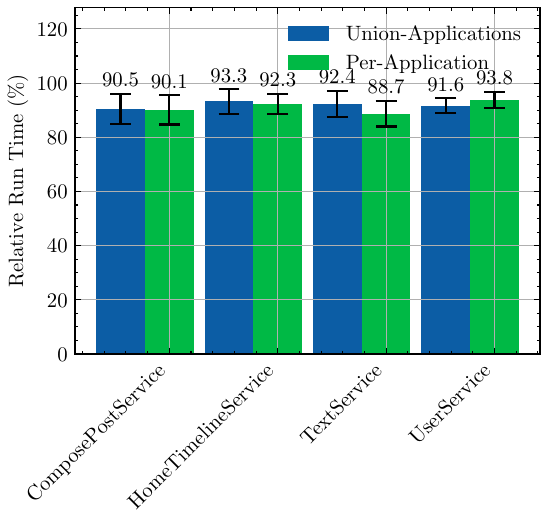}
        \caption{Relative Time}
        \label{fig:all-separate-time}
    \end{subfigure}
    \caption{Comparison of size, memory, and run time of shared libraries reconstructed using per-application inputs versus the union of inputs from all four microservices.}
    \label{fig:cmp-all-separate}
    \vspace{-2em}
\end{figure}

Figure~\ref{fig:cmp-all-separate} compares the relative size, memory usage, and run time of libraries reconstructed using per-application inputs versus those reconstructed using the union of inputs from all four services.
As expected, the shared libraries reconstructed using the union contain more code, resulting in slightly higher relative file sizes and memory usage compared to individually tailored libraries.
Nevertheless, the reductions remain substantial and comparable across both approaches.
Run time behavior shows more variability.
In some cases, the shared version exhibits slightly higher run time, while in others, it performs similarly or better.
This is consistent with earlier observations that run time is influenced by multiple factors, including number of program headers and memory usage.
Overall, these results show that \myapp can reconstruct shared libraries that retain the benefits of reduced size, memory and run time, while still shareable by multiple applications, thereby preserving the key advantage of shared libraries.

\subsection{A Security Case Study: The XZ Backdoor}\label{sec:xz-backdoor}
In our final experiment, we present a case study showing how \myapp handles the XZ backdoor vulnerability, i.e., CVE-2024-3094~\cite{nistCVE20243094}.
XZ is a set of open-source libraries for data compression, which is installed by default on most Linux distributions. Recently, it was discovered that an attacker injected malicious code into a shared library of XZ, \texttt{liblzma.so}~\cite{githubXzutilsBackdoor}.
The malicious code gets loaded into memory when an SSH server is started, allowing the attacker to connect to the SSH server with a private key and execute arbitrary code.
This vulnerability has a CVSS score of 10, indicating the most critical security risk.
Assuming users have installed the compromised \texttt{liblzma.so} library, we investigate whether \myapp can remove the malicious code from the shared library.

To do this, we first reproduce the backdoor~\cite{githubGitHubAmlweemsxzbot}.
We started an SSH server with the compromised \texttt{liblzma.so} library so that the server can be connected in two ways:
(1) using a normal SSH connection, and
(2) using the backdoor connection.
To reconstruct the shared library, we connect to it using a normal SSH connection, which acts as the normal use case.
Then we use \myapp to reconstruct the shared library, using the file ranges collected from the normal SSH connection.
Subsequently, we start the SSH server with the reconstruct library.
We successfully connected to the SSH server using a normal SSH connection, but failed to connect using the backdoor connection.
This indicates that \myapp successfully removed the malicious code from the shared library, preventing the backdoor from being exploited, improving the security.

\section{Discussion}
In this paper, we introduced \myapp, a tool for reconstructing shared libraries based on user-specific needs by retaining only the code that is actually needed.
Our evaluation across nine real-world applications from diverse domains demonstrates that \myapp can reduce file size, memory usage and run time significantly, simultaneously improving security by eliminating unused code and  preserving user-specified functionality of the application.
The reduced size, lower memory usage, and improved memory locality of reconstructed shared libraries make them especially useful for resource-constrained environments like IoT devices and edge computing~\cite{kong2022edge,luo2021resource}.
Additionally, the run time improvements are particularly critical in on-demand computing scenarios—such as serverless functions, where cold start loading time can significantly impact performance~\cite{anup2019agile,li2022help,vahidinia2020cold}.

\textbf{Anticipated Usage}.
\myapp is particularly well-suited for applications with well-defined and narrow usage patterns, such as microservices and serverless functions.
In such contexts,  developers or cloud providers can analyze the usage of shared libraries to determine which portions are actually needed.
This can be achieved through tracing these online services, static analysis, or binary instrumentation, and does not need to be performed in a single pass. Instead, it can be conducted over an extended period in production environments to capture a comprehensive view of usage patterns.
Once sufficient coverage is obtained, \myapp can be used to reconstruct the shared libraries to include only the necessary file ranges, which  leads to performance and security gains.

\textbf{Future Directions}.
Modern software development is largely an additive process: code and dependencies accumulate at every stage, from development to build to deployment.
\myapp introduces a new, subtractive phase that complements this pipeline by removing unused code based on the actual usage of software.
We envision a revised development workflow—develop, build, deploy, and reconstruct—in which reconstruction becomes a standard step to optimize software artifacts post-deployment.
Besides, the performance gains of removing unused code also highlights the extent of inefficiency and code bloat in current systems.
We hope \myapp may drive a broader rethinking of software packaging, modularity, and maintenance.

\section{Conclusion}
We present \myapp, a novel approach for reconstructing shared libraries based on user-specific needs by retaining only the code actually needed.
\myapp takes as input a shared library and a set of file ranges representing the user-needed code segments.
It first cc-out all unneeded code, then applies a two-phase optimization: a merge step that merges nearby file ranges to reduce loading overhead, and a shuffle step that reorders the ranges to minimize file size.
The shuffling problem is formulated as a variant of the Traveling Salesman Problem and solved using a heuristic algorithm.
Finally, \myapp updates the program headers to ensure the shuffled file ranges are correctly mapped to their original memory addresses at runtime.

Our evaluation across nine applications from diverse domains shows that \myapp can reduce shared library sizes by up to 60\%, decrease memory usage by up to 16\%, and improve run time performance by up to 32\%, while preserving functionality and enhancing security through attack surface reduction.

\clearpage
\bibliographystyle{plain}
\bibliography{refs}

\begin{thebibliography}{10}

\bibitem{oracleDebuggingLibrary}
{D}ebugging {L}ibrary ({L}inker and {L}ibraries {G}uide) --- docs.oracle.com.
\newblock \url{https://docs.oracle.com/cd/E19683-01/816-1386/chapter3-33/index.html}.
\newblock [Accessed 29-05-2025].

\bibitem{oracleHeaderLinker}
{E}{L}{F} {H}eader ({L}inker and {L}ibraries {G}uide) --- docs.oracle.com.
\newblock \url{https://docs.oracle.com/cd/E19683-01/816-1386/chapter6-43405/index.html}.
\newblock [Accessed 08-10-2024].

\bibitem{rustlangFeaturesCargo}
{F}eatures - {T}he {C}argo {B}ook --- doc.rust-lang.org.
\newblock \url{https://doc.rust-lang.org/cargo/reference/features.html}.
\newblock [Accessed 26-05-2025].

\bibitem{githubGitHubAmlweemsxzbot}
{G}it{H}ub - amlweems/xzbot: notes, honeypot, and exploit demo for the xz backdoor ({C}{V}{E}-2024-3094) --- github.com.
\newblock \url{https://github.com/amlweems/xzbot}.
\newblock [Accessed 18-01-2025].

\bibitem{man7Ldso8Linux}
ld.so(8) - {L}inux manual page --- man7.org.
\newblock \url{https://man7.org/linux/man-pages/man8/ld.so.8.html}.
\newblock [Accessed 29-05-2025].

\bibitem{nistCVE20243094}
{N}{V}{D} - {C}{V}{E}-2024-3094 --- nvd.nist.gov.
\newblock \url{https://nvd.nist.gov/vuln/detail/CVE-2024-3094}.
\newblock [Accessed 26-09-2024].

\bibitem{githubXzutilsBackdoor}
xz-utils backdoor situation ({C}{V}{E}-2024-3094) --- gist.github.com.
\newblock \url{https://gist.github.com/thesamesam/223949d5a074ebc3dce9ee78baad9e27}.
\newblock [Accessed 18-10-2024].

\bibitem{agadakos2020large}
Ioannis Agadakos, Nicholas Demarinis, Di~Jin, Kent Williams-King, Jearson Alfajardo, Benjamin Shteinfeld, David Williams-King, Vasileios~P Kemerlis, and Georgios Portokalidis.
\newblock Large-scale debloating of binary shared libraries.
\newblock {\em Digital Threats: Research and Practice}, 1(4):1--28, 2020.

\bibitem{agadakos2019nibbler}
Ioannis Agadakos, Di~Jin, David Williams-King, Vasileios~P Kemerlis, and Georgios Portokalidis.
\newblock Nibbler: debloating binary shared libraries.
\newblock In {\em Proceedings of the 35th Annual Computer Security Applications Conference}, pages 70--83, 2019.

\bibitem{ahmad2021trimmer}
Aatira~Anum Ahmad, Abdul~Rafae Noor, Hashim Sharif, Usama Hameed, Shoaib Asif, Mubashir Anwar, Ashish Gehani, Fareed Zaffar, and Junaid~Haroon Siddiqui.
\newblock Trimmer: an automated system for configuration-based software debloating.
\newblock {\em IEEE Transactions on Software Engineering}, 48(9):3485--3505, 2021.

\bibitem{bhattacharya2013combining}
Suparna Bhattacharya, Kanchi Gopinath, and Mangala~Gowri Nanda.
\newblock Combining concern input with program analysis for bloat detection.
\newblock {\em ACM SIGPLAN Notices}, 48(10):745--764, 2013.

\bibitem{biswas2021ancile}
Priyam Biswas, Nathan Burow, and Mathias Payer.
\newblock Code specialization through dynamic feature observation.
\newblock In {\em Proceedings of the Eleventh ACM Conference on Data and Application Security and Privacy}, pages 257--268, 2021.

\bibitem{brown2024broad}
Michael~D Brown, Adam Meily, Brian Fairservice, Akshay Sood, Jonathan Dorn, Eric Kilmer, and Ronald Eytchison.
\newblock A broad comparative evaluation of software debloating tools.
\newblock In {\em 33rd USENIX Security Symposium (USENIX Security 24)}, pages 3927--3943, 2024.

\bibitem{brown2019carve}
Michael~D Brown and Santosh Pande.
\newblock Carve: Practical security-focused software debloating using simple feature set mappings.
\newblock In {\em Proceedings of the 3rd ACM Workshop on Forming an Ecosystem Around Software Transformation}, pages 1--7, 2019.

\bibitem{carvalho2014heartbleed}
Marco Carvalho, Jared DeMott, Richard Ford, and David~A Wheeler.
\newblock Heartbleed 101.
\newblock {\em IEEE security \& privacy}, 12(4):63--67, 2014.

\bibitem{choi2019lambda}
Sean Choi, Muhammad Shahbaz, Balaji Prabhakar, and Mendel Rosenblum.
\newblock $\lambda$-nic: Interactive serverless compute on smartnics.
\newblock In {\em Proceedings of the ACM SIGCOMM 2019 Conference Posters and Demos}, pages 151--152, 2019.

\bibitem{dueck1990threshold}
Gunter Dueck and Tobias Scheuer.
\newblock Threshold accepting: A general purpose optimization algorithm appearing superior to simulated annealing.
\newblock {\em Journal of computational physics}, 90(1):161--175, 1990.

\bibitem{elliott2007anything}
Bill Elliott.
\newblock Anything is possible: Managing feature creep in an innovation rich environment.
\newblock In {\em 2007 IEEE International Engineering Management Conference}, pages 304--307. IEEE, 2007.

\bibitem{fan2020escaping}
Gang Fan, Chengpeng Wang, Rongxin Wu, Xiao Xiao, Qingkai Shi, and Charles Zhang.
\newblock Escaping dependency hell: finding build dependency errors with the unified dependency graph.
\newblock In {\em Proceedings of the 29th ACM SIGSOFT International Symposium on Software Testing and Analysis}, pages 463--474, 2020.

\bibitem{gan2019open}
Yu~Gan, Yanqi Zhang, Dailun Cheng, Ankitha Shetty, Priyal Rathi, Nayan Katarki, Ariana Bruno, Justin Hu, Brian Ritchken, Brendon Jackson, et~al.
\newblock An open-source benchmark suite for microservices and their hardware-software implications for cloud \& edge systems.
\newblock In {\em Proceedings of the Twenty-Fourth International Conference on Architectural Support for Programming Languages and Operating Systems}, pages 3--18, 2019.

\bibitem{gartner2021report}
Gartner.
\newblock Application development will shift to application assembly and integration.
\newblock \url{https://www.gartner.com/en/newsroom/press-releases/2021-11-10-gartner-says-cloud-will-be-the-centerpiece-of-new-digital-experiences}, 2021.
\newblock [Accessed 28-05-2025].

\bibitem{hubert2024bloat}
Bert Hubert.
\newblock Why bloat is still software's biggest vulnerability: A 2024 plea for lean software.
\newblock {\em IEEE Spectrum}, 61(4):22--50, 2024.

\bibitem{junger1995traveling}
Michael J{\"u}nger, Gerhard Reinelt, and Giovanni Rinaldi.
\newblock The traveling salesman problem.
\newblock {\em Handbooks in operations research and management science}, 7:225--330, 1995.

\bibitem{kong2022edge}
Linghe Kong, Jinlin Tan, Junqin Huang, Guihai Chen, Shuaitian Wang, Xi~Jin, Peng Zeng, Muhammad Khan, and Sajal~K Das.
\newblock Edge-computing-driven internet of things: A survey.
\newblock {\em ACM Computing Surveys}, 55(8):1--41, 2022.

\bibitem{koo2019configuration}
Hyungjoon Koo, Seyedhamed Ghavamnia, and Michalis Polychronakis.
\newblock Configuration-driven software debloating.
\newblock In {\em Proceedings of the 12th European Workshop on Systems Security}, pages 1--6, 2019.

\bibitem{li2022help}
Zijun Li, Linsong Guo, Quan Chen, Jiagan Cheng, Chuhao Xu, Deze Zeng, Zhuo Song, Tao Ma, Yong Yang, Chao Li, and Minyi Guo.
\newblock Help rather than recycle: Alleviating cold startup in serverless computing through {Inter-Function} container sharing.
\newblock In {\em 2022 USENIX Annual Technical Conference (USENIX ATC 22)}, pages 69--84, Carlsbad, CA, July 2022. USENIX Association.

\bibitem{liu2023faaslight}
Xuanzhe Liu, Jinfeng Wen, Zhenpeng Chen, Ding Li, Junkai Chen, Yi~Liu, Haoyu Wang, and Xin Jin.
\newblock Faaslight: General application-level cold-start latency optimization for function-as-a-service in serverless computing.
\newblock {\em ACM Transactions on Software Engineering and Methodology}, 32(5):1--29, 2023.

\bibitem{luo2021resource}
Quyuan Luo, Shihong Hu, Changle Li, Guanghui Li, and Weisong Shi.
\newblock Resource scheduling in edge computing: A survey.
\newblock {\em IEEE communications surveys \& tutorials}, 23(4):2131--2165, 2021.

\bibitem{malecha2015occam}
Gregory Malecha, Ashish Gehani, and Natarajan Shankar.
\newblock Automated software winnowing.
\newblock In {\em Proceedings of the 30th Annual ACM Symposium on Applied Computing}, pages 1504--1511, 2015.

\bibitem{anup2019agile}
Anup Mohan, Harshad Sane, Kshitij Doshi, Saikrishna Edupuganti, Naren Nayak, and Vadim Sukhomlinov.
\newblock Agile cold starts for scalable serverless.
\newblock In {\em 11th USENIX Workshop on Hot Topics in Cloud Computing (HotCloud 19)}, Renton, WA, July 2019. USENIX Association.

\bibitem{narasimhan2015copy}
Krishna Narasimhan and Christoph Reichenbach.
\newblock Copy and paste redeemed (t).
\newblock In {\em 2015 30th IEEE/ACM International Conference on Automated Software Engineering (ASE)}, pages 630--640. IEEE, 2015.

\bibitem{navas2023occamv2}
Jorge~A Navas and Ashish Gehani.
\newblock Occam-v2: combining static and dynamic analysis for effective and efficient whole-program specialization.
\newblock {\em Communications of the ACM}, 66(4):40--47, 2023.

\bibitem{paszke2019pytorch}
Adam Paszke, Sam Gross, Francisco Massa, Adam Lerer, James Bradbury, Gregory Chanan, Trevor Killeen, Zeming Lin, Natalia Gimelshein, Luca Antiga, et~al.
\newblock Pytorch: An imperative style, high-performance deep learning library.
\newblock {\em Advances in neural information processing systems}, 32, 2019.

\bibitem{porter2023decker}
Chris Porter, Sharjeel Khan, and Santosh Pande.
\newblock Decker: Attack surface reduction via on-demand code mapping.
\newblock In {\em Proceedings of the 28th ACM International Conference on Architectural Support for Programming Languages and Operating Systems, Volume 2}, pages 192--206, 2023.

\bibitem{porter2020blankit}
Chris Porter, Girish Mururu, Prithayan Barua, and Santosh Pande.
\newblock Blankit library debloating: Getting what you want instead of cutting what you don’t.
\newblock In {\em Proceedings of the 41st ACM SIGPLAN Conference on Programming Language Design and Implementation}, pages 164--180, 2020.

\bibitem{qian2019razor}
Chenxiong Qian, Hong Hu, Mansour Alharthi, Pak~Ho Chung, Taesoo Kim, and Wenke Lee.
\newblock $\{$RAZOR$\}$: A framework for post-deployment software debloating.
\newblock In {\em 28th USENIX security symposium (USENIX Security 19)}, pages 1733--1750, 2019.

\bibitem{quach2019bloat}
Anh Quach and Aravind Prakash.
\newblock Bloat factors and binary specialization.
\newblock In {\em Proceedings of the 3rd ACM Workshop on Forming an Ecosystem Around Software Transformation}, pages 31--38, 2019.

\bibitem{quach2018dpiecewise}
Anh Quach, Aravind Prakash, and Lok Yan.
\newblock Debloating software through $\{$Piece-Wise$\}$ compilation and loading.
\newblock In {\em 27th USENIX security symposium (USENIX Security 18)}, pages 869--886, 2018.

\bibitem{rtrace}
Please refer to the~supplementary material.
\newblock Rtrace: Towards better visibility of shared library execution.
\newblock {\em Anonymized for double-blind review}.

\bibitem{shackleton2023dead}
Will Shackleton, Katriel Cohn-Gordon, Peter~C Rigby, Rui Abreu, James Gill, Nachiappan Nagappan, Karim Nakad, Ioannis Papagiannis, Luke Petre, Giorgi Megreli, et~al.
\newblock Dead code removal at meta: Automatically deleting millions of lines of code and petabytes of deprecated data.
\newblock In {\em Proceedings of the 31st ACM Joint European Software Engineering Conference and Symposium on the Foundations of Software Engineering}, pages 1705--1715, 2023.

\bibitem{sharif2018trimmer}
Hashim Sharif, Muhammad Abubakar, Ashish Gehani, and Fareed Zaffar.
\newblock Trimmer: application specialization for code debloating.
\newblock In {\em Proceedings of the 33rd ACM/IEEE International Conference on Automated Software Engineering}, pages 329--339, 2018.

\bibitem{shen2025understanding}
Yijun Shen, Xiang Gao, Hailong Sun, and Yu~Guo.
\newblock Understanding vulnerabilities in software supply chains.
\newblock {\em Empirical Software Engineering}, 30(1):1--38, 2025.

\bibitem{tariq2025efficient}
Syed Salauddin~Mohammad Tariq, Ali~Al Zein, Soumya~Sripad Vaidya, Arati Khanolkar, Zheng Song, and Probir Roy.
\newblock Efficient serverless cold start: Reducing library loading overhead by profile-guided optimization.
\newblock {\em arXiv preprint arXiv:2504.19283}.
\newblock Note: To appear at ICDCS 2025.

\bibitem{vahidinia2020cold}
Parichehr Vahidinia, Bahar Farahani, and Fereidoon~Shams Aliee.
\newblock Cold start in serverless computing: Current trends and mitigation strategies.
\newblock In {\em 2020 International Conference on Omni-layer Intelligent Systems (COINS)}, pages 1--7, 2020.

\bibitem{xu2014scalable}
Guoqing Xu, Nick Mitchell, Matthew Arnold, Atanas Rountev, Edith Schonberg, and Gary Sevitsky.
\newblock Scalable runtime bloat detection using abstract dynamic slicing.
\newblock {\em ACM Transactions on Software Engineering and Methodology (TOSEM)}, 23(3):1--50, 2014.

\bibitem{zhang2024machine}
Huaifeng Zhang, Mohannad Alhanahnah, Fahmi~Abdulqadir Ahmed, Dyako Fatih, Philipp Leitner, and Ahmed Ali-Eldin.
\newblock Machine learning systems are bloated and vulnerable.
\newblock {\em Proceedings of the ACM on Measurement and Analysis of Computing Systems}, 8(1):1--30, 2024.

\bibitem{zhao2020large}
Nannan Zhao, Vasily Tarasov, Hadeel Albahar, Ali Anwar, Lukas Rupprecht, Dimitrios Skourtis, Arnab~K Paul, Keren Chen, and Ali~R Butt.
\newblock Large-scale analysis of docker images and performance implications for container storage systems.
\newblock {\em IEEE Transactions on Parallel and Distributed Systems}, 32(4):918--930, 2020.

\bibitem{ziegler2019honey}
Andreas Ziegler, Julian Geus, Bernhard Heinloth, Timo H{\"o}nig, and Daniel Lohmann.
\newblock Honey, i shrunk the elfs: Lightweight binary tailoring of shared libraries.
\newblock {\em ACM Transactions on Embedded Computing Systems (TECS)}, 18(5s):1--23, 2019.

\end{thebibliography}

\clearpage
\appendix
\section{Appendix}
\subsection{The Absolute Values of $g$ and Corresponding Percentiles}
See Table~\ref{tab:g-values} for the absolute values of $g$ and corresponding percentiles for the applications evaluated in this paper.
\begin{table*}
    \centering
    \scalebox{0.8}{
        \begin{tabular}{lrrrrrrrrrrrrrrrrrrr}
            \toprule
            test\_name            & 0-th & 10-th & 20-th & 30-th & 40-th & 50-th & 60-th & 70-th & 80-th & 90-th & 91-th & 92-th & 93-th  & 94-th  & 95-th  & 96-th  & 97-th  & 98-th  & 99-th  \\
            \midrule
            ComposePostService    & 0    & 74    & 154   & 282   & 442   & 674   & 1114  & 1753  & 3866  & 14466 & 15420 & 16580 & 24854  & 28477  & 31614  & 40333  & 44584  & 60609  & 86708  \\
            HomeTimelineService   & 0    & 74    & 154   & 282   & 467   & 714   & 1114  & 1840  & 3814  & 14509 & 15479 & 16602 & 24893  & 28488  & 31651  & 40411  & 44641  & 60726  & 86887  \\
            TextService           & 0    & 74    & 154   & 266   & 435   & 666   & 1114  & 1741  & 3789  & 14293 & 15288 & 16493 & 24697  & 25839  & 31463  & 40025  & 44445  & 60144  & 85992  \\
            UserService           & 0    & 74    & 138   & 234   & 394   & 594   & 1050  & 1678  & 3930  & 12370 & 14562 & 15651 & 18694  & 23192  & 26450  & 30396  & 35497  & 47131  & 74470  \\
            imagemagick           & 0    & 218   & 442   & 954   & 1610  & 2394  & 3754  & 7978  & 15514 & 32106 & 35306 & 39290 & 45402  & 46602  & 52010  & 59018  & 78890  & 131066 & 162730 \\
            inference\_cnn\_mnist & 0    & 122   & 378   & 805   & 1482  & 3186  & 6810  & 15066 & 32586 & 75369 & 83767 & 94883 & 109719 & 128421 & 153494 & 199209 & 270481 & 449687 & 749020 \\
            nmap                  & 0    & 74    & 378   & 618   & 1194  & 1706  & 3626  & 8298  & 16170 & 50778 & 54464 & 61974 & 74647  & 92571  & 114586 & 151053 & 247516 & 353814 & 591450 \\
            popler                & 0    & 58    & 122   & 186   & 282   & 514   & 944   & 1667  & 3757  & 20400 & 23206 & 27600 & 36759  & 42040  & 48510  & 63500  & 76966  & 109632 & 169661 \\
            train\_cnn\_mnist     & 0    & 138   & 442   & 941   & 1802  & 3914  & 7882  & 15961 & 32278 & 70646 & 80269 & 93719 & 105133 & 124140 & 147784 & 185752 & 266465 & 440391 & 748621 \\
            \bottomrule
        \end{tabular}
    }
    \caption{Absolute values of $g$ and corresponding percentiles for the applications. The values are in bytes.}
    \label{tab:g-values}
\end{table*}

\subsection{The Congruence Requirement}
According to the ELF spefications, the file offset and memory address where the file offset is loaded should meet the congruence constraint:
\begin{equation}
    \label{eq:congruence}
    \texttt{file\_offset} \equiv \texttt{memory\_address} \pmod{P}
\end{equation}
In other words, the file offset and the memory address should have the same modulo of the page size $P$.

Given a list of file ranges $R=\{r_1, r_2,\dots,r_n\}$ their loaded memory addresses, the start address of each file range is denoted $m_i$.
Since the original shared library must meet the congruence constraint, we have, for any $i$, $r_i.s  \equiv m_i \pmod{P}$.
Since we remap the shuffled ranges to their original memory addresses, we also have $r'_i.s \equiv m_i \pmod{P}$, where $r_i'$ is the shuffled file range.
Therefore, we have the congruence constraint for the shuffled file ranges as shown in Equation~\ref{eq:congruence-shuffled}:
\begin{equation*}
    r'_i.s \equiv r_i.s \pmod{P}
\end{equation*}

\subsection{The Range Overlapping Problem}

When the system program loader loads a shared library, it follows the program headers, mapping each specified file range into memory.
For each range, the loader performs a page-aligned mapping: it rounds down the starting offset to the nearest page boundary ("page down") and rounds up the end offset ("page up").
Consequently, the entire page(s) covering the file range are loaded into memory.

Consider three original file ranges $R=\{r_1, r_2, r_3\}$ where $r_1$ resides in page $p_1$, and $r_2$, $r_3$ reside in page $p_2$.
And $r_1.s \equiv r_2.s \pmod{P}$, where $P$ is the page size.
If \tspshuffle reorders these ranges and moves $r_3$ to page $p_1$, a problem arises.
If  the loader loads $r_1$, then $r_2$, then $r_3$.
When $r_1$ is loaded, it maps to its original memory address with no issue.
The same applies to $r_2$.
However, when $r_3$ is loaded into page $p_2$, it triggers a page-aligned mapping, which includes all of file content in page $p_1$.
Thereby unintentionally overwriting $r_2$ with the content of $r_1$ in memory, this corrupts the memory content, and will cause wrong behavior when the program accesses $r_2$ later.
We refer to this issue as the \emph{range overlapping problem}.

This issue only occurs when the relative order of file ranges is changed.
As \linearshuffle preserves the relative order, it avoids this problem.
In contrast, \tspshuffle can introduce it because it may change the relative order of file ranges.

\end{document}